\documentclass[runningheads]{llncs}
\usepackage[T1]{fontenc}
\usepackage{graphicx}
\usepackage{algpseudocode}
\usepackage{algorithm}
\usepackage{amsmath}
\usepackage{amssymb}
\usepackage{amsfonts}
\usepackage{latexsym}
\usepackage{bbm}
\usepackage{bm}
\usepackage{makecell}
\usepackage{booktabs}
\usepackage{multirow}
\usepackage{hyperref}
\usepackage{marvosym}

\usepackage{color}

\hypersetup{
    colorlinks=true,
    citecolor=blue,
    linkcolor=blue,
    urlcolor=blue
}
\begin{document}
\title{Quantifying Knee Cartilage Shape and Lesion: From Image to Metrics}
%
%
\author{
Yongcheng Yao \Letter \inst{1} \and
Weitian Chen\inst{2}
}

%
\authorrunning{Y. Yao and W. Chen}
%
\institute{
School of Informatics, The University of Edinburgh, United Kingdom\\
\email{yc.yao@ed.ac.uk} \and
CU Lab of AI in Radiology (CLAIR), Department of Imaging and Interventional Radiology, The Chinese University of Hong Kong, Hong Kong\\
\email{wtchen@cuhk.edu.hk}
}
\maketitle              
\begin{abstract}
Imaging features of knee articular cartilage have been shown to be potential imaging biomarkers for knee osteoarthritis. Despite recent methodological advancements in image analysis techniques like image segmentation, registration, and domain-specific image computing algorithms, only a few works focus on building fully automated pipelines for imaging feature extraction. In this study, we developed a deep-learning-based medical image analysis application for knee cartilage morphometrics, CartiMorph Toolbox (CMT). We proposed a 2-stage joint template learning and registration network, CMT-reg. We trained the model using the OAI-ZIB dataset and assessed its performance in template-to-image registration. The CMT-reg demonstrated competitive results compared to other state-of-the-art models. We integrated the proposed model into an automated pipeline for the quantification of cartilage shape and lesion (full-thickness cartilage loss, specifically). The toolbox provides a comprehensive, user-friendly solution for medical image analysis and data visualization. The software and models are available at \url{https://github.com/YongchengYAO/CMT-AMAI24paper}.

\end{abstract}

\section{Introduction}
\label{sec:intro}
Osteoarthritis (OA) represents a predominant cause of disability among the elderly demographic, with knee OA being a highly prevalent sub-type of the disease. In light of the global trend towards an aging population, an increase in the burden of OA can be expected. The pathogenesis of knee OA involves deterioration in various knee joint tissues including the articular cartilage, subchondral bone, meniscus, and ligament. Notably, the loss of articular cartilage integrity represents a hallmark feature of knee OA progression. Cartilage morphometrics is a promising tool for deriving biomarkers from magnetic resonance imaging (MRI) or radiograph. The combination of medical image acquisition, image computing, and visualization has the potential to facilitate knee OA monitoring, pathology research, treatment evaluation, and surgical planning. Despite the advances in semiautomated cartilage lesion quantification  \cite{wirth2008technique,williams2010anatomically,maerz2016surface,favre2017anatomically}, automated lesion grading \cite{guermazi2017brief,dorio2020association}, and knee joint tissue segmentation \cite{xu2019deepatlas,gaj2020automated,khan2022deep,li2024source}, a fully automated image analysis application is rarely encountered. To this end, we developed CartiMorph Toolbox (CMT) for automatic shape and lesion analysis. It is a deep-learning-powered cartilage morphometrics solution, from an image to quantitative metrics.

\textbf{Contribution.} CMT consists of deep learning (DL) models for tissue segmentation and registration. We proposed a method for joint template learning and registration, CMT-reg. We implemented and improved the cartilage morphometrics framework, CartiMorph \cite{yao2024cartimorph}, and developed a integrated suite of modules for computing environment configuration, project management, DL model life-cycle management (including training, fine-tuning, evaluation, and inference), and data visualization. The software and models are publicly available \footnote{\url{https://github.com/YongchengYAO/CMT-AMAI24paper}}.

\section{Related Work}
\label{sec:related-work}

\textbf{Deep Segmentation Model.} Deep learning has been attracting increasing interesting in medical image segmentation due to its superior performance and learning ability. Advancements in DL-based knee tissue segmentation include the combination of convolutional neural network (CNN) with shape model \cite{liu2018deep,ambellan2019automated}, alpha matte \cite{khan2022deep}, and self attention \cite{liang2022position}. There is increasing number of medical image segmentation models based on vision Transformer in recent years \cite{liu2022isegformer,li2023sdmt}. Most deep segmentation models are trained and optimized for specific tasks. It is desired to have DL models that generalize well on out-of-distribution unseen data. The auto-configuring nnU-Net \cite{isensee2018nnu} and the foundation model MedSAM \cite{ma2024segment} demonstrate significant generalisability, offering a strong foundation for advanced deep learning applications in medical imaging.  We implemented the 3D variant of nnU-Net in CMT.

\textbf{Deep Registration Model.} VoxelMorph \cite{balakrishnan2019voxelmorph} represents a successful application of deep learning for medical image registration. Knee image registration models have been developed \cite{xu2019deepatlas,shen2019networks,ding2022aladdin}. We compared CMT-reg with 2 state-of-the-art (SOTA) algorithms: (i) LapIRN \cite{mok2020large}, the SOTA method in Learn2Reg \cite{hering2022learn2reg} challenge, and (ii) Aladdin \cite{ding2022aladdin}, which achieves SOTA performance in atlas-as-a-bridge alignment.

\textbf{Cartilage Shape \& Lesion Analysis.} Shape and lesion analysis of knee cartilage includes the quantification of healthy anatomy and lesion. Methods for cartilage thickness mapping \cite{maerz2016surface,favre2017anatomically}, semiautomated lesion quantification \cite{wirth2009regional}, and subregional parcellation \cite{panfilov2022deep,favre2017anatomically} have been developed. CartiMorph \cite{yao2024cartimorph} integrated deep segmentation and registration models in the cartilage morphometrics pipeline, and achieved improvement in thickness mapping, parcellation, and full-thickness cartilage loss (FCL) estimation. In this work, we proposed methods to simplify the CartiMorph framework and yet achieve superior performance in template-to-image registration. Specifically, we used a single CNN for joint template learning and image registration.

\section{Methods}
\label{sec:methods}

\subsection{Toolbox Overview}
\label{subsec:toolbox-overview}

\begin{figure}
\centering
\includegraphics[width=\textwidth]{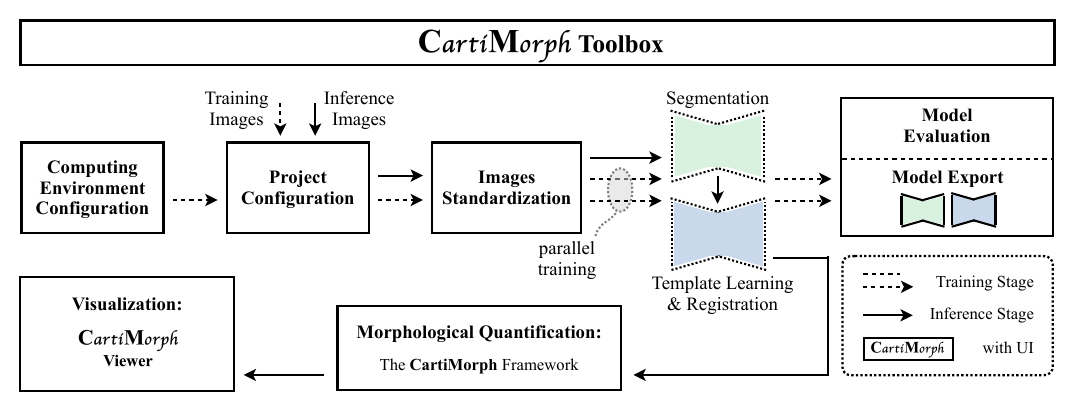}
\caption{Structure of CartiMorph Toolbox (CMT).} 
\label{fig1}
\end{figure}

CMT comprises of modules for computing environment configuration, project configuration, image standardization, DL models management, morphological quantification, and data visualization (Fig.~\ref{fig1}). The configurations of data source, data analysis steps, hyper-parameters setting, and figures and metrics visualization are integrated in CMT user interface. The toolbox was developed to address several challenges in achieving fully automated DL applications for cartilage morphometrics.

\textbf{DL Models Life-cycle Management.} The development of DL models involves the configuration of model training environment, data processing, and model evaluation. A user-friendly mechanism for model sharing and fine-tuning is useful for adapting pre-trained models to new domains. A plugin-and-use model inference tool may facilitate the development of DL applications. CMT management the whole life-cycle of DL models (in the toolbox) by providing user interface for models training, fine-tuning, evaluation, inference, and sharing.

\textbf{Image Standardization.} This module consists of image intensity normalization, image re-orientation, and image resampling. It is an important data pre-processing step for DL model training and downstream algorithms. The standard image orientation in CMT is RAS+ where the image dimensions correspond to the right, anterior, and superior directions.

\textbf{Sub-regional Parcellation.} CartiMorph \cite{yao2024cartimorph} introduced approaches for rule-based cartilage parcellation and regional FLC estimation, which have only been validated on right-side knee images. More importantly, deviation from the standard scanning position could lead to inaccurate partitioning results. We enhanced the parcellation method to accommodate both left and right knee images by flipping the left knee images before and after the algorithm. The robustness against scanning position variation is improved by rigid-body registration to the learned template image.

\subsection{DL Models}
\label{subsec:DL-models}

\begin{figure}
\centering
\includegraphics[width=0.9\textwidth]{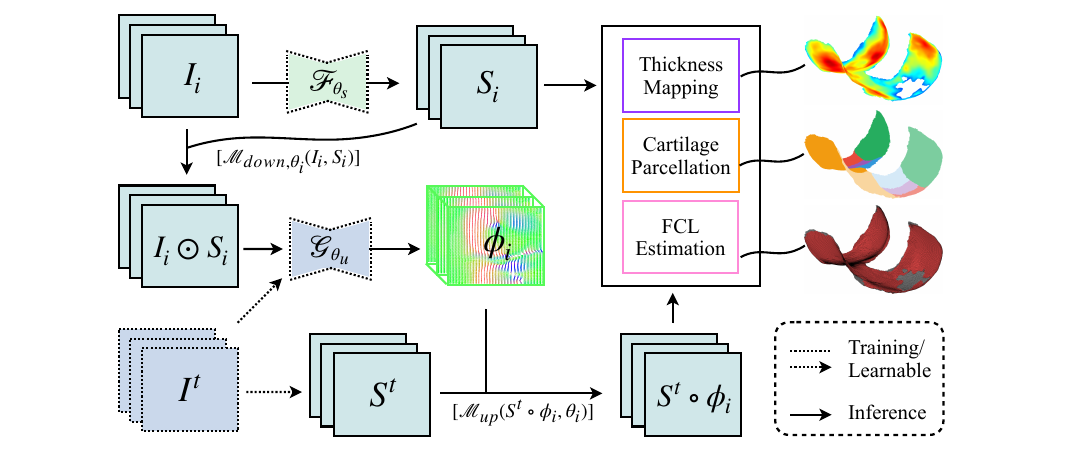}
\caption{Deep learning models and quantification algorithms in CMT.}
\label{fig2}
\end{figure}

Fig.~\ref{fig2} provides an overview of the 2 DL models and image computing algorithms in CMT. A segmentation CNN $F_{\boldsymbol{\theta}_s}$ is trained on $\{ (\boldsymbol{I}_i, \boldsymbol{S}^l_i)\}_{i=1}^{n_\text{t}}$, where $\boldsymbol{I}_i$ is the input image and $\boldsymbol{S}^l_i$ is the ground truth segmentation label. We integrate the template learning and image registration into a single CNN $G_{\{\boldsymbol{\theta}_u,\boldsymbol{I}^t\}}$, which is trained on the masked and optionally down-sampled image $\boldsymbol{I}_i \odot \boldsymbol{S}^l_i$. The down-sampling step aims to accommodate the model on a GPU with limited memory by leveraging spatial redundancy in the dense deformation field $\boldsymbol{\phi}_i$. A template image $\boldsymbol{I}^t$ is learned from $G_{\{\boldsymbol{\theta}_u,\boldsymbol{I}^t\}}$ and the corresponding template segmentation mask $\boldsymbol{S}^t$ is constructed via registration. Specifically, we register all training images into the template space and apply the image-to-template deformation field to their ground truth mask; a probability map $\boldsymbol{P}^t$ can therefore be calculated as the average warped mask
\begin{equation}
\label{eq:temp-1}
     [\boldsymbol{\phi}^{-1}_i], \boldsymbol{\phi}_i = G_{\{\boldsymbol{\theta}_u,\boldsymbol{I}^t\}}((\boldsymbol{I}_i \odot \boldsymbol{S}^l_i)),
\end{equation}
\begin{equation}
\label{eq:temp-2}
     \boldsymbol{P}^t = \frac{1}{n_\text{t}} \sum_{i=1}^{n_\text{t}}  (\boldsymbol{S}^l_i \circ \boldsymbol{\phi}^{-1}_i),
\end{equation}
where $\boldsymbol{\phi}^{-1}_i$ and $\boldsymbol{\phi}_i$ denotes image-to-template and template-to-image deformation field, respectively; finally, the template mask $\boldsymbol{S}^t$ is obtained by thresholding $\boldsymbol{P}^t$. 

Once $F_{\boldsymbol{\theta}_s}$ and $G_{\{\boldsymbol{\theta}_u,\boldsymbol{I}^t\}}$ finish training, the predicted segmentation mask $\boldsymbol{S}_i$ and the deformation field $\boldsymbol{\phi}_i$ are utilized in subsequent algorithms for cartilage thickness mapping, cartilage parcellation, and FCL estimation. We adopted 3D nnU-Net as the segmentation model and the proposed CMT-reg as the template-learning-and-registration model.

\subsection{Joint Template Learning and Registration Model}
\label{subsec:model-archi}

\begin{figure}
\centering
\includegraphics[width=0.8\textwidth]{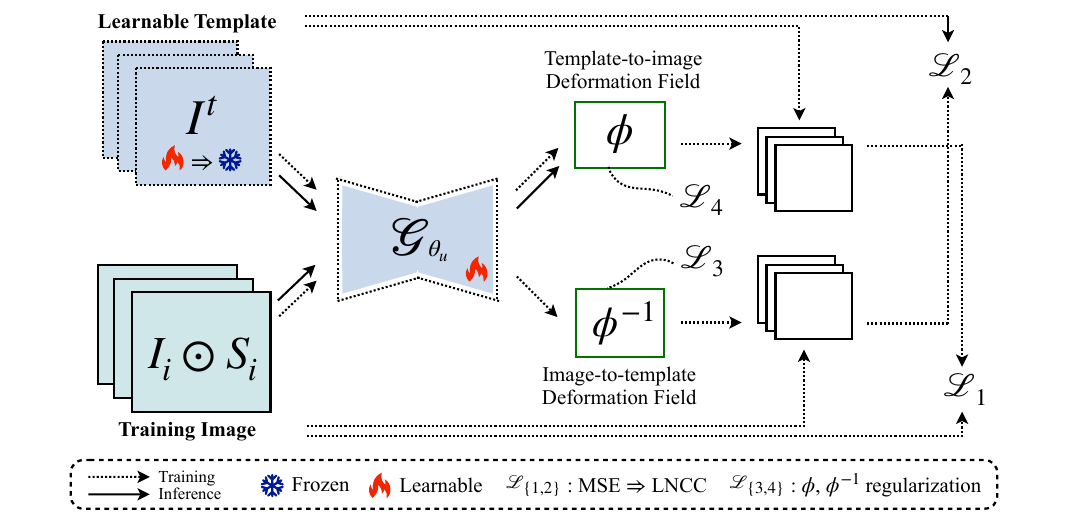}
\caption{The proposed joint template learning and registration model, CMT-reg.}
\label{fig3}
\end{figure}

Template construction, involving the creation of a representative image from multiple aligned images, is intrinsically tied to image registration. With the advances of deep registration models, methods for template learning have been developed. Common image similarity losses include mean squared error (MSE), cross-correlation (CC), normalized cross-correlation (NCC), local normalized cross-correlation (LNCC). Previous studies \cite{balakrishnan2019voxelmorph,yao2024cartimorph} have demonstrated that LNCC is superior to MSE in image registration task. However, the influence of different loss functions on template image quality remains under-explored. We present the results of template images trained with various losses in Section \ref{sec:experiments-results}.

Based on our observation that the LNCC loss excels in image registration and MSE in template learning, we propose to jointly learn the template and train the registration CNN in a 2-stage strategy (Fig.~\ref{fig3}). In the first training stage, both the template and CNN are learnable. In the second stage, the template is frozen while the CNN are trainable. The model is optimized with image similarity losses $\mathcal{L}_{\{1,2\}}$ and deformation field regularization $\mathcal{L}_{\{3,4\}}$. For the image similarity loss, MSE is used in stage 1 and LNCC is used in stage 2.

\section{Experiments \& Results}
\label{sec:experiments-results}

\subsection{Registration Model Evaluation}
In CMT, the FCL estimation algorithm depends on the alignment between the template mask $\boldsymbol{S}^t$ and the model predicted segmentation mask $\boldsymbol{S}_i$. Combined with mesh processing algorithms, the warped template mask $\boldsymbol{S}^t \circ \boldsymbol{\phi}_i$ is used to reconstruct the intact cartilage surface. The ideal behavior of the registration model is that the warped template mask could cover as much FCL regions as possible without encompassing non-cartilage tissues.

\textbf{Metrics.} We first adopted the conventional evaluation metrics, the Dice similarity coefficient (DSC) and the 95th percentile Hausdorff distance (HD95), to quantify the discrepancy between $\boldsymbol{S}^t \circ \boldsymbol{\phi}_i$ and $\boldsymbol{S}_i$. Additionally, we measured the relative difference between the surface area of the bone-cartilage interface from $\boldsymbol{S}^t \circ \boldsymbol{\phi}_i$ and that of the manually calibrated pseudo-healthy bone-cartilage interface $\boldsymbol{M}_i^{\text{pseudo}}$: $(A_i-A_i^{\text{pseudo}})/A_i^{\text{pseudo}}$, $A_i^{\text{pseudo}}=\mathcal{O}_{\text{surfArea}}(\boldsymbol{M}_i^{\text{pseudo}})$, where $A_i=\mathcal{O}_{\text{surfArea}}(\mathcal{O}_{\text{surfSeg}}(\boldsymbol{S}^t \circ \boldsymbol{\phi}_i))$, and $\mathcal{O}_{\text{surfSeg}}(\cdot)$ generates the bone-cartilage interface mesh from a 3D mask, $\mathcal{O}_{\text{surfArea}}(\cdot)$ calculates surface area.

\textbf{Models.} We trained variants of LapIRN and Aladdin. Like CMT-reg, Aladdin is a joint template learning and registration model, while LapIRN is solely an image registration model. For LapIRN, we first trained models and used an external template, \emph{CLAIR-Knee-103R}, from the CartiMorph study\footnote{Data from \url{https://github.com/YongchengYAO/CartiMorph}} for template-to-image evaluation. Two variants of LapIRN were trained: (i) LapIRN-diff, a model formulated with the stationary velocity field, and (ii) LapIRN-disp, a model formulated with displacement field. We trained variants of Aladdin under various setting of image similarity loss (MSE, NCC, and LNCC) and evaluation target ($\mathcal{L}_{sim}$: ordinal image similarity loss; $\mathcal{L}^i_{p}$: pair similarity loss in image space; $\mathcal{L}^t_{p}$: pair similarity loss in template space).

\textbf{Implementation Details.} CMT requires conda for managing virtual environments. Our implementation of nnU-Net (v1.7.0) was released as a python package, CartiMorph-nnUNet. The proposed CMT-reg is based on VoxelMorph and was released as CartiMorph-vxm. Models were trained with NVIDIA A100 (80G). A window size of 27 was used in LNCC loss for all models. The channel numbers of CMT-reg is $\{[48,96,96,96]^{\text{enc}},[96]^{\text{neck}},[96,96,96,96]^{\text{dec}},[48,48]^{\text{conv}}\}$, corresponding to the encoder, bottleneck, decoder, and the size-preserving convolutional block, respectively.

\subsection{Data}
We evaluated the toolbox with the OAI-ZIB dataset that consists of 507 MR images and manual segmentation labels for the femoral cartilage (FC), tibial cartilage (TC), femur, and tibia. We subdivided the tibial cartilage into the medial tibial cartilage (MTC) and lateral tibial cartilage (LTC). Data split is shown in Table~\ref{tab:data}, where the same testing set is reserved for model evaluation. The CMT-reg and LapIRN were trained with data split 1, while Aladdin models were trained with data split 2. The Kellgren–Lawrence (KL) grades were employed in stratified random sampling to guarantee the inclusion of cases with varying cartilage lesion severities in both the training and testing sets. Subset 3 includes 12 cases from the testing set, featuring severe cartilage lesions, was utilized to evaluate the models' behavior based on the relative surface area difference.

\setlength{\tabcolsep}{5pt}
\begin{table}
\caption{Datasets for model training, testing, and analysis. Dataset 1 and 2 share the same testing set. Dataset 3 is a subset of the testing set, used for analyzing the behavior of registration models.}
\label{tab:data}
\centering
\begin{tabular}{cccc}
    \hline
    \textbf{Data Split} & \textbf{Train (\%)} & \textbf{Validation (\%)}  & \textbf{Test (\%)[KL Grade: 0-4, N/A]} \\
    \hline 
    $\#1$ &  404 (80\%) & 0 & \multirow{2}{*}{103 (20\%) [21,12,22,28,15,5]} \\
    $\#2$ &  324 (64\%) & 80 (16\%)  & {}   \\
    \hline
    \textbf{Subset} & \multicolumn{3}{c}{\textbf{Analysis [KL Grade: 0-4, N/A]}} \\
    \hline
    $\#3$ & \multicolumn{3}{c}{12 [0,0,0,6,6,0]} \\
    
    \hline
\end{tabular}
\end{table}

\subsection{Results \& Discussion}

\begin{figure}[h!]
\centering
\includegraphics[width=\textwidth]{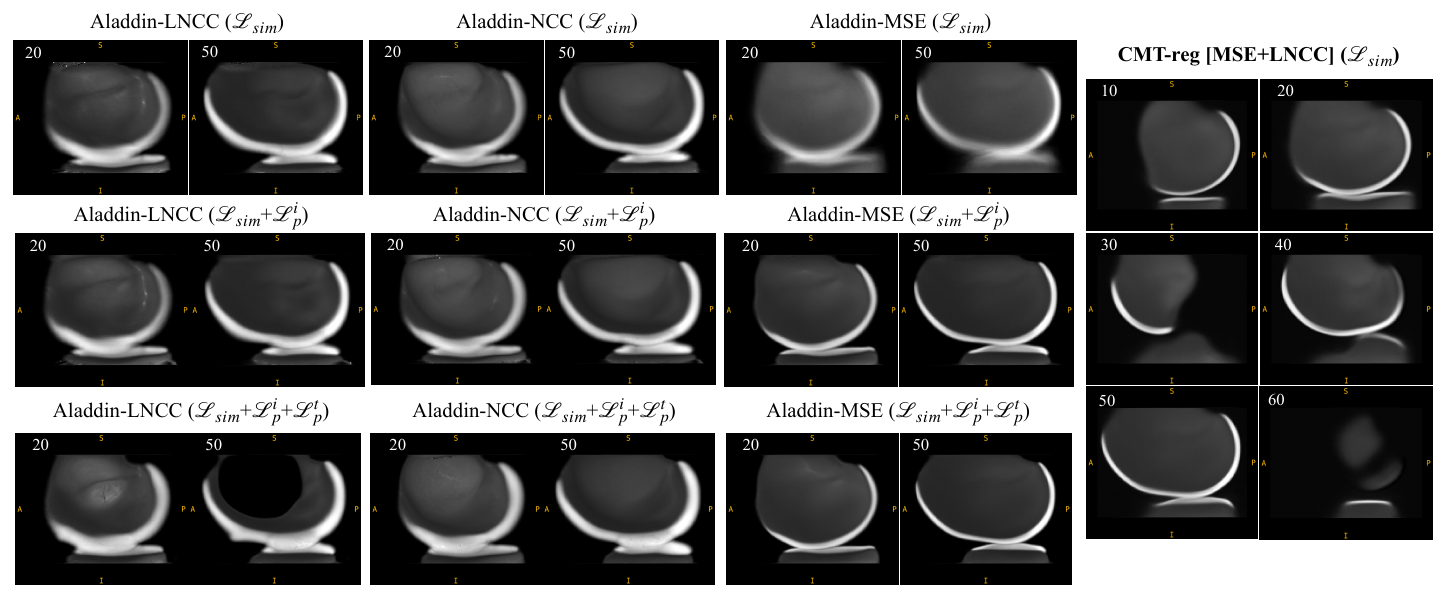}
\caption{Learned template images from variants of Aladdin and the proposed model, CMT-reg. Numbers in the upper left corner are indices of sagittal slices.}
\label{fig4}
\end{figure}

\textbf{Template Image Quality.} Fig.~\ref{fig4} shows that template images learned from models with MSE loss exhibit greater sharpness compared to those produced with LNCC loss. Additionally, models trained with LNCC loss sometimes produced corrupted images, such as the Aladdin-LNCC ($\mathcal{L}_{sim}$+$\mathcal{L}^i_{p}$) model in Fig.~\ref{fig4}. This observation informed our model training strategy, wherein we used MSE loss to learn the template image and then switch to LNCC loss for continuing the registration network training. Although Aladdin models achieve high registration accuracy with noisy and blurry template images, we prefer a model that learns a high-quality template, as it can be used in general registration models like LapIRN.

\textbf{Balanced Registration Models.} Table~\ref{tab2} shows that our model is comparable to SOTA models in terms of volume overlap (DSC) and surface distance (HD95). In our application, the desired registration model must achieve a balance between accuracy (evaluated by DSC and HD95) and FCL region coverage (evaluated by relative surface area difference). Table~\ref{tab3} shows our model outperforms Aladdin models and is comparable to LapIRN models in FCL region coverage.

\setlength{\tabcolsep}{5pt}
\begin{table}[t!]
\caption{Template-to-image registration accuracy.}
\label{tab2}
\centering
\begin{tabular}{ccccccc}
    \hline
    \multirow{2}{*}{\textbf{Model (Loss)}} & \multicolumn{3}{c}{\textbf{DSC $\uparrow$}} & \multicolumn{3}{c}{\textbf{HD95 $\downarrow$ (mm)}} \\
    \cmidrule(lr){2-4} \cmidrule(lr){5-7} 
    {} & \textbf{FC} & \textbf{mTC} & \textbf{lTC} & \textbf{FC} & \textbf{mTC} & \textbf{lTC} \\
    \hline
    Aladdin-LNCC ($\mathcal{L}_{sim}$) & \textbf{0.905} & 0.816 & 0.839 & \textbf{0.58} & \textbf{0.92} & 0.94\\
    Aladdin-LNCC ($\mathcal{L}_{sim}$+$\mathcal{L}^i_{p}$) & \textbf{0.905} & 0.817 & 0.839 & 0.59 & \textbf{0.92} & 0.92 \\
    Aladdin-LNCC ($\mathcal{L}_{sim}$+$\mathcal{L}^i_{p}$+$\mathcal{L}^t_{p}$) & 0.904 & 0.809 & 0.831 & 0.64 & 1.09 & 1.16 \\
    Aladdin-NCC ($\mathcal{L}_{sim}$) & 0.904 & 0.818 & 0.842 & 0.59 & \textbf{0.92} & 0.89 \\
    Aladdin-NCC ($\mathcal{L}_{sim}$+$\mathcal{L}^i_{p}$) & \textbf{0.908} & \textbf{0.829} & \textbf{0.858} & \textbf{0.55} & \textbf{0.85} & \textbf{0.72} \\
    Aladdin-NCC ($\mathcal{L}_{sim}$+$\mathcal{L}^i_{p}$+$\mathcal{L}^t_{p}$) & 0.904 & 0.819 & 0.840 & 0.62 & 0.99 & 1.03 \\
    Aladdin-MSE ($\mathcal{L}_{sim}$) & 0.505 & 0.368 & 0.305 & 3.49 & 5.69 & 6.51 \\
    Aladdin-MSE ($\mathcal{L}_{sim}$+$\mathcal{L}^i_{p}$) & 0.889 & 0.808 & 0.849 & 0.65 & 0.97 & \textbf{0.76} \\
    Aladdin-MSE ($\mathcal{L}_{sim}$+$\mathcal{L}^i_{p}$+$\mathcal{L}^t_{p}$) & 0.885 & 0.803 & 0.841 & 0.65 & 0.98 & 0.82 \\
    \hline
    LapIRN-diff ($\mathcal{L}_{sim}$) & 0.855 & 0.740 & 0.731 & 1.09 & 1.69 & 2.01 \\
    LapIRN-disp ($\mathcal{L}_{sim}$) & 0.898 & 0.806 & \textbf{0.850} & 0.81 & 1.38 & 1.28 \\
    \hline
    CMT-reg ($\mathcal{L}_{sim}$) & 0.895 & \textbf{0.821} & \textbf{0.850} & 0.70 & 1.10 & 0.89 \\
    \hline
\end{tabular}
\end{table}

\setlength{\tabcolsep}{5pt}
\begin{table}[t!]
\caption{Registration models behavior analysis.}
\label{tab3}
\centering
\begin{tabular}{cccc}
    \hline
    \multirow{2}{*}{\textbf{Model (Loss)}} & \multicolumn{3}{c}{\textbf{ $(A_i-A_i^{\text{pseudo}})/A_i^{\text{pseudo}} \; \mid \downarrow \mid$}} \\
    \cmidrule(lr){2-4}
    {} & \textbf{FC} & \textbf{mTC} & \textbf{lTC} \\
    \hline
    Aladdin-LNCC ($\mathcal{L}_{sim}$) & -0.035 & -0.115 & -0.140 \\
    Aladdin-LNCC ($\mathcal{L}_{sim}$+$\mathcal{L}^i_{p}$) & -0.029 & -0.106 & -0.143 \\
    Aladdin-LNCC ($\mathcal{L}_{sim}$+$\mathcal{L}^i_{p}$+$\mathcal{L}^t_{p}$) & -0.026 & -0.080 & -0.133 \\
    Aladdin-NCC ($\mathcal{L}_{sim}$) & -0.031 & -0.112 & -0.150 \\
    Aladdin-NCC ($\mathcal{L}_{sim}$+$\mathcal{L}^i_{p}$) & -0.029 & -0.098 & -0.145 \\
    Aladdin-NCC ($\mathcal{L}_{sim}$+$\mathcal{L}^i_{p}$+$\mathcal{L}^t_{p}$) & -0.032 & -0.088 & -0.134 \\
    Aladdin-MSE ($\mathcal{L}_{sim}$) & -0.203 & -0.393 & -0.468 \\
    Aladdin-MSE ($\mathcal{L}_{sim}$+$\mathcal{L}^i_{p}$) & -0.021 & -0.079 & -0.142 \\
    Aladdin-MSE ($\mathcal{L}_{sim}$+$\mathcal{L}^i_{p}$+$\mathcal{L}^t_{p}$) & \textbf{-0.002} & -0.070 & -0.129 \\
    \hline
    LapIRN-diff ($\mathcal{L}_{sim}$) & \textbf{0.007} & -0.087 & -0.134 \\
    LapIRN-disp ($\mathcal{L}_{sim}$) & 0.040 & \textbf{-0.047} & \textbf{-0.103} \\
    \hline
    CMT-reg ($\mathcal{L}_{sim}$) & 0.036 & \textbf{-0.048} & \textbf{-0.124} \\
    \hline
\end{tabular}
\end{table}

\begin{figure}[t!]
\centering
\includegraphics[width=\textwidth]{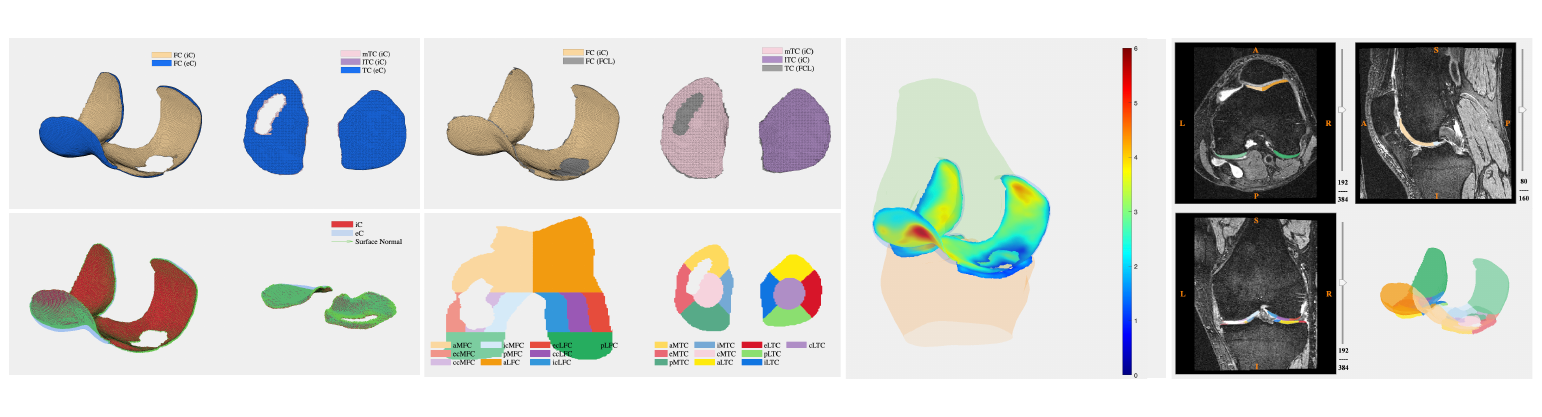}
\caption{Examples of data visualization in CartiMorph Viewer (CMV)}
\label{fig5}
\end{figure}

\textbf{Data Visualization.} We showcased examples of figures from the visualization module, CartiMorph Viewer (CMV), in Fig.~\ref{fig5}. Additionally, CMV displays regional metrics (including volume, thickness, surface area, and FCL), facilitating the direct correlation of qualitative measurements with visual representations.

\section{Conclusion}
\label{sec:conclusion}
In this study, we developed the CartiMorph Toolbox (CMT), a medical AI application for knee cartilage morphometrics. The key component, CMT-reg, is a 2-stage joint template learning and registration network. Trained on the public OAI-ZIB dataset, CMT-reg demonstrated satisfying balance between accuracy and FCL region coverage. CMT offers an out-of-the-box solution for medical image computing and data visualization. 

\textbf{Prospect of application.} The toolbox offers an AI solution for medical image computing, enabling precise quantification of cartilage shape and lesion. It automates the complicated process of model training, inference, and image analysis. Its potential applications include clinical diagnostics support system, treatment planning, and research on knee cartilage health.

\begin{credits}
\subsubsection{\ackname} This work was supported by the United Kingdom Research and Innovation (grant EP/S02431X/1), UKRI Centre for Doctoral Training in Biomedical AI at the University of Edinburgh, School of Informatics. For the purpose of open access, the author has applied a creative commons attribution (CC BY) licence to any author accepted manuscript version arising.

\subsubsection{\discintname}
The authors have no competing interests to declare that are
relevant to the content of this article.
\end{credits}

%
\bibliographystyle{splncs04}
\bibliography{Paper-0041}

\end{document}